\documentclass[12pt]{iopart}
\usepackage{epsfig,harvard}

\begin{document}
\paper[(e,2e) on lithium]{Spin-resolved electron-impact ionization of lithium}
\author{Igor Bray\footnote{electronic address: I.Bray@flinders.edu.au},
Justin Beck and Chris Plottke
}
\address{
Electronic Structure of Materials Centre,
The Flinders University of South Australia,
G.P.O. Box 2100, Adelaide 5001, Australia}
\date{\today}

\begin{abstract}
Electron-impact ionization of lithium is studied using the convergent
close-coupling (CCC) method at 25.4 and 54.4~eV. Particular attention
is paid to the spin-dependence of the ionization cross
sections. Convergence is found to be more rapid for the spin
asymmetries than for the
underlying cross sections.  Comparison with the recently measured and
DS3C-calculated data of \citeasnoun{Streun99} is most
intriguing. Excellent agreement is found with the measured and
calculated spin asymmetries, yet the discrepancy between the CCC and
DS3C cross sections is very large.
\end{abstract}
\pacs{34.80.Bm, 34.80.Dp}
\jl{2}
\submitted
\maketitle

\section{Introduction}
One of the great strengths of the field of electron-atom scattering is 
the strong interplay between theory and experiment. Generally,
experiment is used to test theory and so leads the way forward. The
ultimate goal of a measurement of a particular scattering process is
to perform a so-called complete experiment, one that fully determines
all possible aspects of the collision \cite{B69a,B69b}. Such
measurements are able to fully test the scattering amplitudes arising
in any theory. In practice this goal has been rarely achieved, with
electron-impact excitation of the $n$P states of helium being a
notable example (see \citeasnoun{FB95} and references therein). However,
this scattering process is not particularly 
rich in information requiring the measurement, at each scattering angle,
of only the differential 
cross section and the three Stokes parameters to fully test the
two independent theoretical (complex) scattering amplitudes. One way
to increase the scattering 
process information is to scatter electrons from a non-spin-zero
atomic target. This immediately doubles the number of scattering
amplitudes, and allows for a thorough test of the theoretical
treatment of electron-exchange processes. It was electron scattering
on the H and Na targets which showed the accuracy of the convergent
close-coupling (CCC) approach to electron-atom
scattering \cite{BS93l,B94}.

Another way to increase the amount of information associated with a
scattering process is to consider excitation of more complicated final 
states or, at energies above the ionization threshold, to
measure the fully resolved differential ionization (e,2e) cross
sections. Though initially the close-coupling approach has been
applied to elastic and discrete excitation processes, it has been
clear for some time that intermediate excitation of the continuum must 
be treated in order to ensure accurate final results. The CCC method
has proved particularly successful in this regard and showed that
e-Na spin-resolved 3P excitation description required
detailed coupling within the continuum~\cite{B94R}. This unexpected
result encouraged us to look directly at ionization processes.

The CCC application to electron-impact ionization of helium has been
particularly successful 
 for asymmetric \cite{BF96,REBFM96,REBFM96l} and
equal-energy sharing
\cite{BFRE97_64,BFRE98,Rioual98} kinematics. Some problems with absolute values
were identified \cite{REBF97} and explained by reference to the
mechanics of the close-coupling method \cite{B97l}. While the total
ionization cross section (TICS) is very stable at all energies as a function
of the number of 
states $N$ used to expand the total wave function, the underlying singly
differential ionization cross section (SDCS) may not be. The TICS is
stable due to the unitarity of the formalism, which does not allow for
any double-counting of the ionization cross sections even though these 
are obtained from excitation of pseudostates of positive energy
$0<\epsilon_n^{(N)}<E$ where $E$ is the total (excess) energy. We
believe the instability in the SDCS is due 
to convergence to zero of excitation amplitudes of states with
$E/2<\epsilon_n^{(N)}<E$. This leads to a step-function SDCS.
Though this
explanation has been met with some hostility \cite{BC99}, who claim
that the CCC-calculated SDCS should instead be symmetric about $E/2$
(at odds with unitarity!) we are still
confident that our interpretation is correct \cite{B99reply}, and enhanced
by the work of \citeasnoun{S99l}. Whereas we
supposed that the CCC amplitudes at the step would converge extremely
slowly in magnitude to the full size of the step, \citeasnoun{S99l} showed, in
effect, that the CCC amplitudes have converged, but to half the step
size. His analysis reconciles the coherent versus incoherent
combination of the amplitudes at $E/2$, see discussion by \citeasnoun{B99jpb}.
Furthermore, other approaches have 
shown consistency and accuracy of the CCC method
\cite{MKW99,BRIM99}. 

What we find more disturbing is that a systematic 
application of the CCC method to e-H ionization has shown inconsistent 
agreement with experiment \cite{B99jpb}. In the energy region between
17.6 and 30~eV there is at times substantial disagreement with
experiment, not found in the e-He case. 
The e-H and e-He ionization problems differ in that the former needs
to be solved for two total spins.  This may cause the CCC method more
difficulty in treating targets like H as opposed to He.
Having spin-resolved fully differential e-H ionization
measurements would be particularly helpful. Unfortunately, this has
not yet occurred, but there has been substantial progress in measuring
closely related e-Li ionization processes
\cite{Betal92,Streun98,Streun99}.

The aim of this paper is to examine the implications of the analysis
of \citeasnoun{S99l} on spin-resolved cross sections in the case of
54.4~eV e-Li ionization. In addition, we compare the CCC method with the recent
measurements and calculations of e-Li ionization at 25.4~eV by
\citeasnoun{Streun99}. 

\section{Theory}
The details of the CCC theory for ionization of atoms by electron
impact have been given by \citeasnoun{BF96}. This was suitable for
hydrogen and helium targets with extension to lithium being trivial
using the details given by \citeasnoun{B94}. Briefly, the structure of 
the lithium atom is obtained by diagonalising the target frozen-core
Hartree-Fock Hamiltonian $H_2$ using an orthogonal (one-electron) Laguerre
basis. This is possible after the core 1s wave function has been
evaluated using the self-consistent field Hartree-Fock equations. Upon 
diagonalisation $n=1,\dots,N$ square-integrable target states $\phi_n^{(N)}$
with energies $\epsilon_n^{(N)}$ are obtained. The
negative-energy states describe the discrete spectrum, while the
positive-energy states provide a discretization of the target
continuum. For a given incident electron energy $k_i^2/2$ the
close-coupling equations are solved for the $T$ matrix, separately for
each total spin $S$,
to define the scattering amplitudes for excitation of states
$\phi_f^{(N)}$ of energy $\epsilon_f^{(N)}<E$
\begin{equation}
f_S^{(N)}(\bi{k}_f,\bi{k}_i)\equiv \langle
\bi{k}_f(1)\phi_f^{(N)}(2)|T_S|\phi_i^{(N)}(2)\bi{k}_i(1)\rangle,
\label{damp}
\end{equation}
where $E=k_i^2/2+\epsilon_i^{(N)}$ is the total (excess) energy, and
where the numbers in parenthesis indicate electron space.

Thus, the problem looks as if only discrete excitation has been
treated. We associate ionization with excitation of the
positive-energy states. Assuming that the asymptotic Hamiltonian may
be partitioned asymmetrically as $K_1+H_2$, where $K_1$ is the
projectile-space kinetic energy operator, leads to the
definition of the (e,2e) amplitude
\begin{equation}
f_S^{(N)}(\bi{k}_f,\bi{q}_f,\bi{k}_i)\equiv 
\langle \bi{q}_f^{(-)}|\phi_f^{(N)}\rangle\langle
\bi{k}_f\phi_f^{(N)}|T_S|\phi_i^{(N)}\bi{k}_i\rangle,
\label{rawamp}
\end{equation}
where $\langle \bi{q}_f^{(-)}|$ is a continuum eigenstate of $H_2$ of
energy $q_f^2/2=\epsilon_f^{(N)}$  \cite{BF96}.

The fundamental problem with the application of the close-coupling
approach to ionization is that allowance for excitation of states with 
$0<\epsilon_f^{(N)}<E$, equivalent to integration over the continuum
from $0$ to $E$,  suggests double-counting of ionization
processes. Yet being a unitary theory, which yields accurate total
ionization cross sections (TICS) \cite{BS93l}, is a contradiction to
this. The step-function idea \cite{B97l}, suggested upon numerical
investigation, says that with increasing $N$ the amplitudes for
excitation of states with $E/2<\epsilon_f^{(N)}<E$ will converge to
zero, with the resulting secondary energy integration being from $0$
to $E/2$ as in formal theory of ionization. The question remains as to 
how to define the cross sections for finite $N$. The incoherent
combination of amplitudes on either side of $E/2$, 
\begin{equation}
\frac{d^3\sigma_S^{(N)}}{d\Omega_1 d\Omega_2 dE_2}=
|f_S^{(N)}(\bi{k}_f,\bi{q}_f,\bi{k}_i)|^2+
|f_S^{(N)}(\bi{q}_f,\bi{k}_f,\bi{k}_i)|^2,
\label{old}
\end{equation}
was suggested by
\citeasnoun{BF96} as a way of preserving unitarity.
This combination had nothing to do with Pauli symmetrization, only
aimed at
preserving the accuracy of the TICS for the energy integration ending
at $E/2$. The individual $f_S$ are already
a coherent combination of their own direct and exchange
amplitudes. The second term
must not be confused with an ``exchange'' term. It is a numerical
``double-counting'' term which should be near zero for $q_f<k_f$.
Typically, the term with $q_f<k_f$ is by far the most
dominant. Only at $E/2$ are the two terms quite similar. 

While \eref{old}  yielded excellent angular distributions the
magnitude of the $E/2$ cross sections was typically around a factor of two
too low \cite{BFRE97_64,BFRE98,Rioual98}. We put this down to
extremely slow convergence of amplitudes to the top of the step size.
\citeasnoun{S99l} analysed the close-coupling approach to ionization
in a model e-H problem and concluded that at $E/2$ the cross section
should be defined as 
\begin{equation}
\frac{d^3\sigma_S^{(N)}}{d\Omega_1 d\Omega_2 dE_2}=
|f_S^{(N)}(\bi{k}_f,\bi{q}_f,\bi{k}_i)+
(-1)^Sf_S^{(N)}(\bi{q}_f,\bi{k}_f,\bi{k}_i)|^2,
\label{new}
\end{equation}
and that the CCC amplitudes at $E/2$ had converged but to half the
step size, as if solving the CCC equations is like performing
Fourier expansions of step functions. Hence \eref{old} should be
multiplied by exactly two, but only for the case of equal
energy-sharing, and hence not affecting the accuracy to which unitarity is
satisfied. For asymmetric energy sharing \citeasnoun{S99l} showed that 
only if the CCC amplitudes were identically zero in the energy range 
$E/2<\epsilon_f^{(N)}<E$ could the ionization amplitudes be
unambiguously defined. In which case they are just the CCC amplitudes
in the energy region $0<\epsilon_f^{(N)}\le E/2$. For practical
purposes he suggested that the error with using \eref{new} generally
would be relatively small.

At first glance one might expect substantial difference between
\eref{new} and \eref{old}. For substantially asymmetric energy sharing 
$q_f<k_f$ only the first term contributes significantly, and so both
prescriptions give a similar result. At $E/2$ we need to multiply
\eref{old} by two before comparison with \eref{new}.  A detailed
numerical comparison for e-H ionization has been given
\cite{B99jpb} and found that
the two forms were barely distinguishable. The reason for this is
that if $q_f=k_f$ then
\begin{equation}
f_S^{(N)}(\bi{k}_f,\bi{q}_f,\bi{k}_i)=
(-1)^Sf_S^{(N)}(\bi{q}_f,\bi{k}_f,\bi{k}_i)+
\delta_S^{(N)}(\bi{k}_f,\bi{q}_f,\bi{k}_i),
\label{approx}
\end{equation}
where $\delta_S^{(N)}$ is a relatively small number in practical
calculations, which should converge to zero with increasing $N$. The
difference between $2\times$\eref{old} and \eref{new} is
$|\delta_S^{(N)}|^2$. This reconciles the two approaches and yields
similar results in realistic e-H calculations.
We should mention that the problem with lack of convergence in the
magnitudes of the CCC amplitudes at asymmetric energy-sharing remains
and we resort to a semi-empirical rescaling of these amplitudes
utilising the known values at $E/2$ \cite{B99jpb}. 

\section{Results}
\begin{figure}
\epsffile{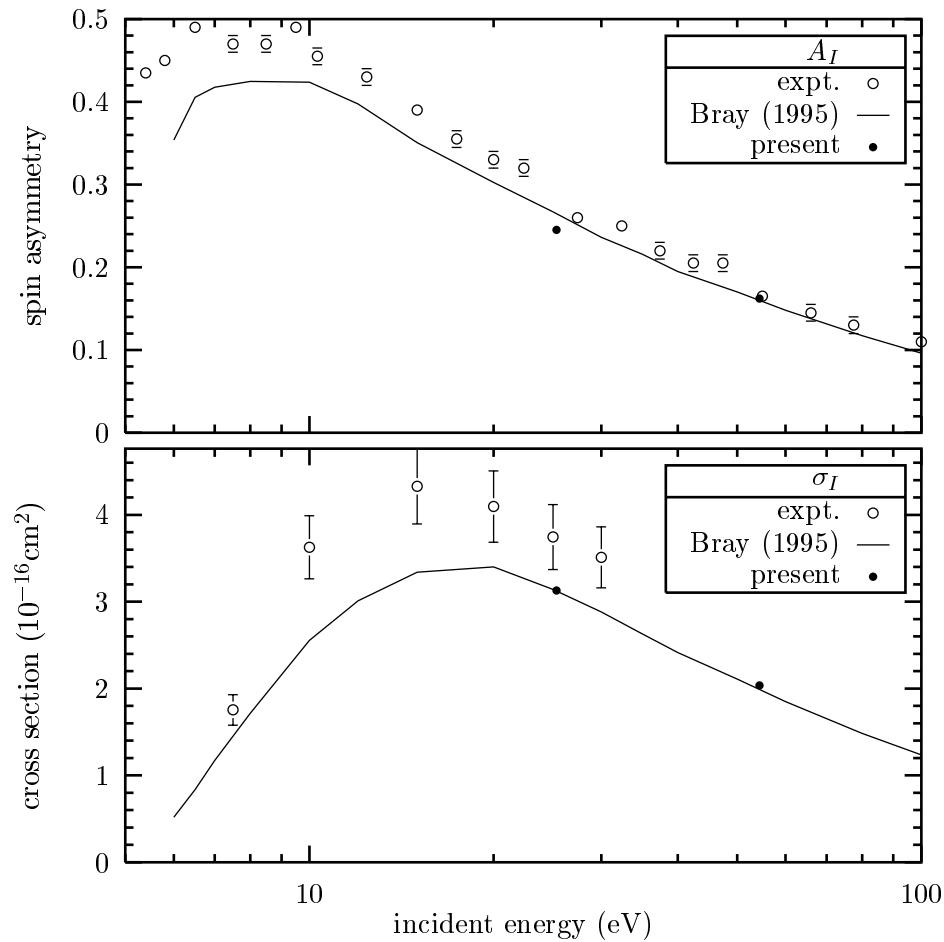}
\caption{Total ionization cross sections and spin asymmetries for
electron-impact ionization of the ground state of lithium. The results 
of the two present 
calculations are indicated by the solid dots. The spin asymmetry
experimental data are due  
to \protect\citeasnoun{BMRS85} and the total ionization data are due
to \protect\citeasnoun{ZA69}. The solid curve is due to CCC calculations 
reported by \protect\citeasnoun{B95jpbl}.}
\label{tics}
\end{figure}

Before looking at the detailed results for the two incident electron
energies considered  
we present, in \fref{tics}, the total ionization cross sections and
their spin asymmetries. These are compared with previous, much
smaller, CCC calculations and experiment. Good agreement is found of
the present calculations with the old for both parameters. Agreement
with the spin asymmetry measurements is satisfactory given the stated
systematic uncertainty \cite{BMRS85}. Convergence for the cross
section is very good, and thus the systematic discrepancy with the
measurements of \citeasnoun{ZA69} persists.
\subsection{Electron-lithium ionization at 54.4 eV}
Experimental data for e-Li ionization is available for 54.4 eV
incident energy ($E=49$~eV), with secondary electrons having energy
(eV) sharing as
$(E_B,E_A)$ pairs of (5,44), (14,35) and (24.5,24.5) \cite{Streun98}. 
A CCC(69) calculation has already been presented for these cases by
\citeasnoun{Streun98}. Here we also give a CCC(97) calculation which
has maximum target-space orbital angular momentum $l_{\rm max}=6$ with 
around 18-$l$ states for each $l$. The 
69-state calculation has $l_{\rm max}=5$ with around 15-$l$ states for 
each $l$. The 97-state calculation
requires approximately 1.5~G of RAM. In addition to checking
convergence in both the spin asymmetries and differential cross
sections, for the (24.5,24.5) case we also consider the generation of
these using both the 
combinations \eref{old} multiplied by two and \eref{new}.

\begin{figure}
\epsffile{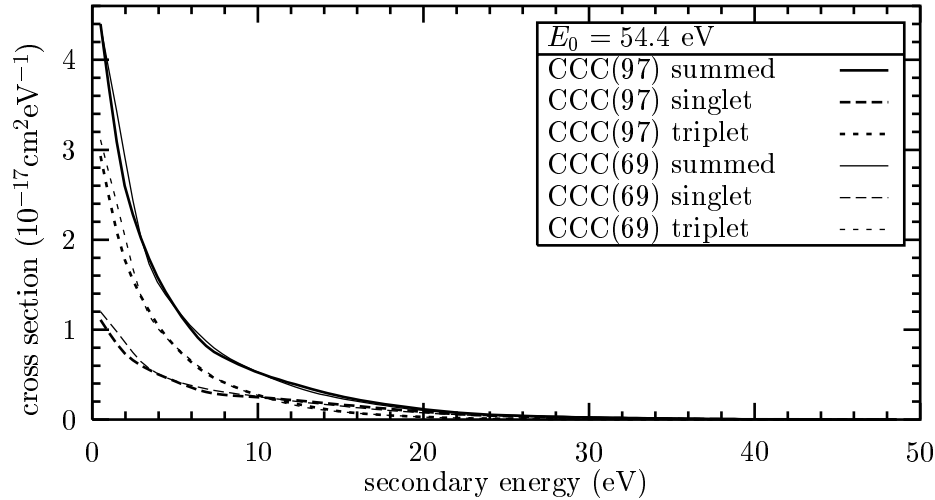}
\caption{The singly differential cross section for 54.4~eV
electron-impact ionization of the ground state of lithium calculated
using the 69- and 97-state approximations. The singlet and triplet
cross sections include the spin weights.}
\label{54.4_sdcs}
\end{figure}

In \fref{54.4_sdcs} we compare the SDCS arising in the 69- and
97-state calculations. We see that both the singlet and triplet
components are rather smooth and fall-off to zero uniformly. In
experiment the SDCS would be measured to be symmetric about 24.5~eV,
and the present results should not be viewed as a contradiction to
this, as \citeasnoun{BC99} have. What is plotted is the square of the
magnitude of the amplitude
\eref{rawamp} integrated over the angles of the outgoing electrons. If 
we were to use the form \eref{old} or \eref{new} an almost identical
symmetric SDCS would result, with the difference that \eref{new} would 
yield a cross section two times bigger at the secondary energy of
24.5~eV. The plotted SDCS yield TICS upon secondary energy integration 
from $0$ to $E$, whereas the symmetric forms would have $E/2$ as the
endpoint of integration.
It is the smallness of the cross section at 24.5~eV that, in
our view, allows the CCC method to yield relatively accurate SDCS at
the smaller secondary energies. In other words, the step-function is
relatively easy to satisfy for the 54.4~eV incident energy, and all of 
the possible physical processes are treated by the pseudostates with
$\epsilon_f^{(N)}\le E/2$.

\begin{figure}
\epsffile{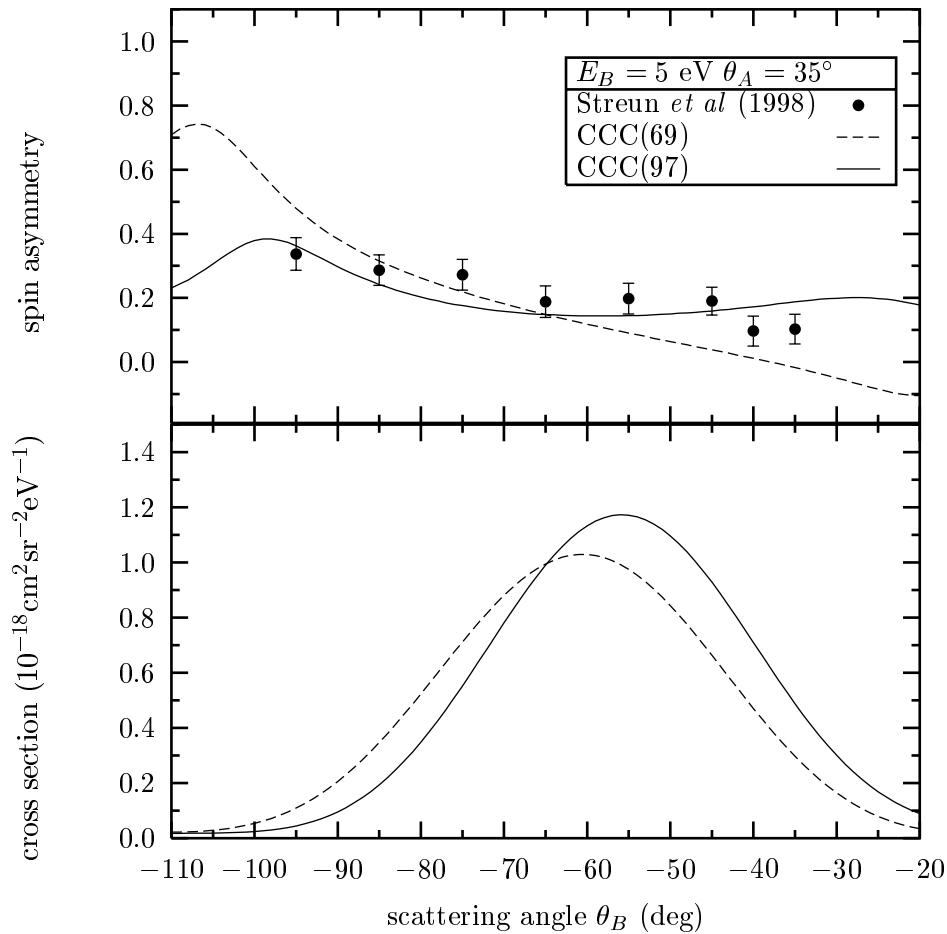}
\caption{The coplanar $\theta_A=35^\circ$, $E_B=5$~eV triply differential cross
section and spin asymmetry for 54.4~eV 
electron-impact ionization of the ground state of lithium calculated
using the 69- and 97-state approximations. The measurements and the
CCC(69) calculation have been reported by \protect\citeasnoun{Streun98}.}
\label{54.4_5}
\end{figure}

The spin-averaged triply differential cross sections (TDCS) is
\begin{equation}
\frac{d^3\sigma^{(N)}}{d\Omega_1 d\Omega_2 dE_2}=
\frac{1}{4}\frac{d^3\sigma_{S=0}^{(N)}}{d\Omega_1 d\Omega_2 dE_2}+
\frac{3}{4}\frac{d^3\sigma_{S=1}^{(N)}}{d\Omega_1 d\Omega_2 dE_2},
\label{tdcs}
\end{equation}
and the corresponding spin asymmetry is
\begin{equation}
A=(1-r)/(1+3r),
\label{asym}
\end{equation}
where $r$ is the ratio of the triplet to singlet (no spin weights) TDCS.
These are given for the secondary energy of $E_B=5$~eV in
\fref{54.4_5} in the
geometry where the fast (44~eV) electron is detected at
$\theta_A=35^\circ$. We use the notation, convenient in the coplanar
geometry, that negative angles are on the opposite side of the
incident beam ($z$-axis) to the positive angles.
We see that there is some difference between the two calculations with 
the asymmetry varying more where the cross section is small. This is an 
example of the general statement that the smaller the cross section
the bigger the calculation necessary to obtain such cross sections
accurately. Further, even larger calculations, which grow very rapidly
with increasing $l_{\rm max}$, would be necessary to determine the
cross sections more accurately. Nevertheless agreement with experiment 
is satisfactory for both the spin asymmetries, and is superior to
the distorted wave Born approximation (DWBA) reported by
\protect\citeasnoun{Streun98}. 

\begin{figure}
\epsffile{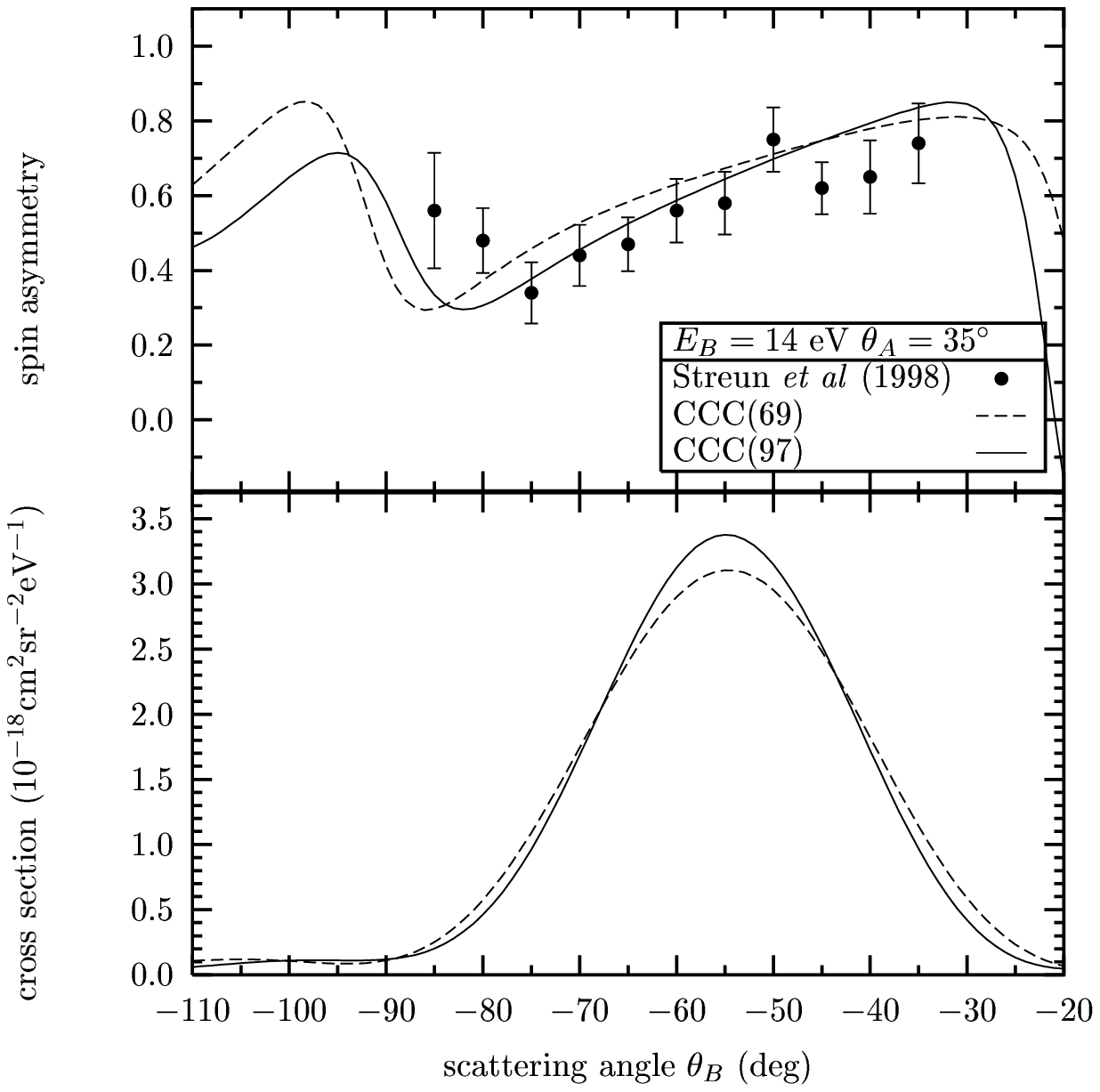}
\caption{The coplanar $\theta_A=35^\circ$, $E_B=14$~eV  triply
differential cross 
section and spin asymmetry for 54.4~eV 
electron-impact ionization of the ground state of lithium calculated
using the 69- and 97-state approximations. The measurements and the
CCC(69) calculation have been reported by \protect\citeasnoun{Streun98}.}
\label{54.4_14}
\end{figure}

\Fref{54.4_14} reports results for the $E_B=14$~eV case. Somewhat
better convergence is found for this case for both the spin
asymmetries and the TDCS. Agreement with experiment is good and is of
the same quality to that of the DWBA calculation reported by
\protect\citeasnoun{Streun98}.  

\begin{figure}
\epsffile{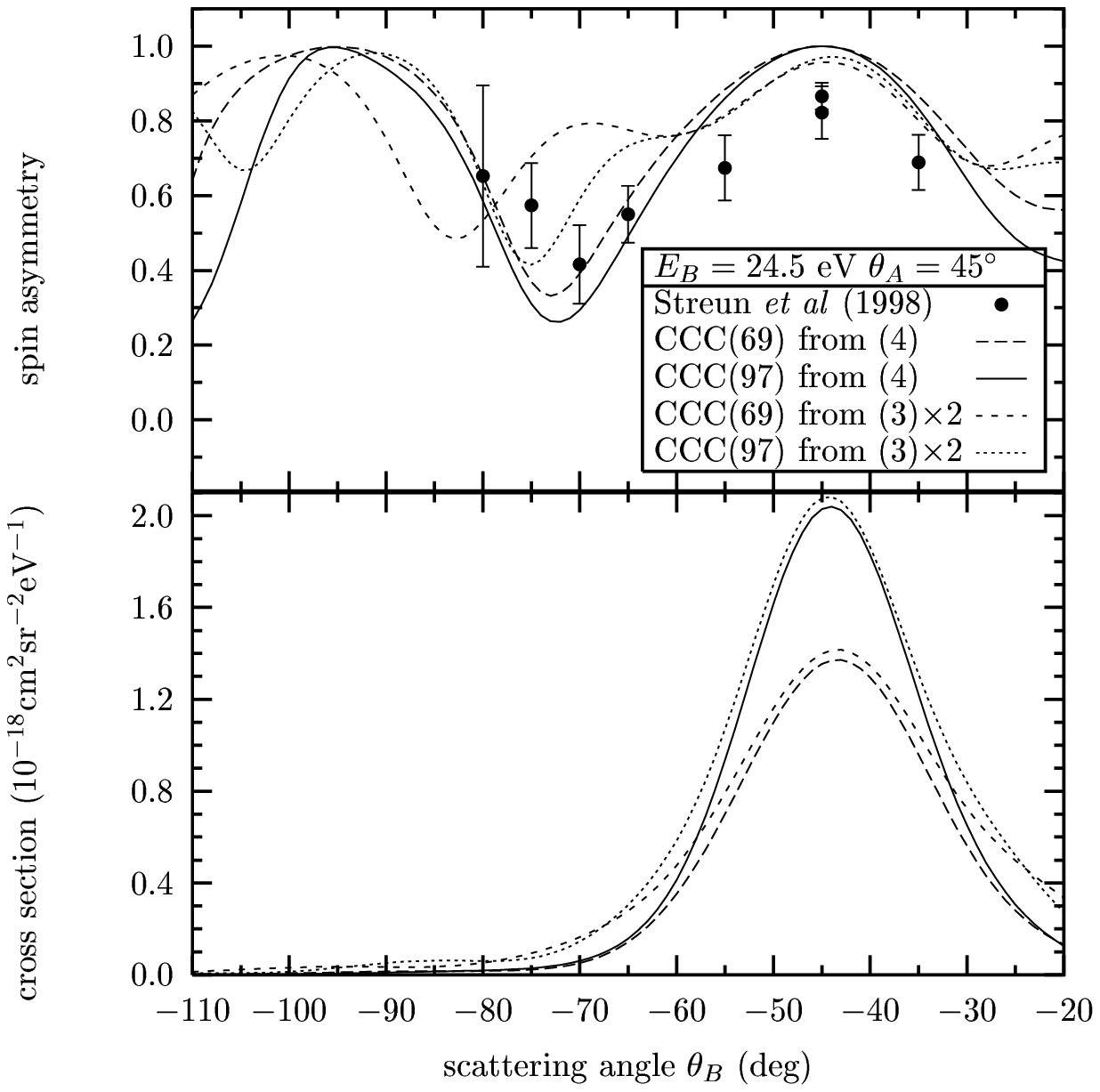}
\caption{The coplanar $\theta_A=45^\circ$, $E_B=24.5$~eV  triply
differential cross 
section and spin asymmetry for 54.4~eV 
electron-impact ionization of the ground state of lithium calculated
using the 69- and 97-state approximations. The cross sections and the spin
asymmetries have been calculated using \eref{old} and \eref{new}.
The measurements and the
CCC(69) calculation have been reported by \protect\citeasnoun{Streun98}.}
\label{54.4_24.5}
\end{figure}

The equal-energy sharing case $E_B=E_A=24.5$~eV presented in
\fref{54.4_24.5} is more
interesting. For the two previous cases combinations
\eref{old} and \eref{new} yield near identical results due to the second
``double-counting'' term being negligible, see \fref{54.4_sdcs} for
energies greater than 24.5~eV. However, at equal energy-sharing both
terms come into play as they 
are derived from the same amplitudes \eref{rawamp}. Due to the
suggestion of \citeasnoun{S99l} that at $E/2$ the CCC amplitudes
converge to half the step size, the results of \eref{old} need to be
multiplied by two before comparison with \eref{new}, which only affects 
the TDCS and not the spin asymmetries.

Let us examine the TDCS first. Around the maximum of the TDCS both
forms give much the same result, which differ substantially for the two 
calculations. The difference between the CCC(69) and CCC(97) curves is 
not surprising, and is primarily due to the increase by one of $l_{\rm 
max}$ in the latter calculation. Given that the outgoing electron
energies are 24.5~eV one would expect to require even substantially
larger than $l_{\rm max}=6$. However, the unitarity of the
close-coupling formalism severely restricts flux to large angular
momentum channels as this is related to convergence in just the
elastic scattering channel \cite{B94l}. The agreement between the two
combinations  
of amplitudes for the larger TDCS indicates that $\delta_S^{(N)}$ in
\eref{approx} is relatively small at those angles. Thus, we see that
while satisfying \eref{approx} is a good thing, the difference between the
CCC(97) and CCC(69) TDCS indicates that it is not sufficient to assure accuracy
of the results. Clearly calculations with even larger $l_{\rm max}$
are necessary to obtain a more accurate estimate of the TDCS.

Turning our attention to the spin asymmetries, we see that the form
\eref{new} yields identically unity for the
$\theta_A=-\theta_B=45^\circ$ case. This is expected from the Pauli
Principle with the triplet TDCS vanishing identically in \eref{new}
and almost so in \eref{old} due to \eref{approx}. What is particularly 
interesting is that \eref{new} yield much closer results for the two
CCC calculations than does \eref{old}. Most encouraging is the
removal, by the use of  \eref{new} instead of  \eref{old},
of the oscillation around $\theta_B=-70^\circ$, not seen in experiment 
or the DWBA theory \cite{Streun98}.  This means that the combination
\eref{new} is more 
efficient at hiding a lack of convergence in the underlying CCC
amplitudes by the construction of the exactly required symmetry in the
amplitudes used to generate the TDCS.

\subsection{Electron-lithium ionization at 25.4 eV}
The recent measurements of \citeasnoun{Streun99} at 25.4~eV incident
energy ($E=20$~eV) take a different form to those above. The coplanar symmetric
$\theta_A=-\theta_B=45^\circ$ single point is taken, and the
asymmetry measured at this point as a function of the energy-sharing of 
the two electrons. Given the difficulty the CCC theory has in
obtaining convergent SDCS at low energies, such measurements are
particularly challenging for the theory.

We present the results of just a single CCC calculation, which has
been checked for convergence. It couples a
total of 107 states, where $l_{max}=7$ and has around 18-$l$ states
for each $l$. The Laguerre exponential fall-offs  $\lambda_l\approx2$
were varied slightly to ensure a pseudostate of 10~eV for each
$l$ resulting with a total of around ten other positive-energy states. The
amplitudes for arbitrary $E_B$ are obtained by interpolation \cite{BF96}.

\begin{figure}
\epsffile{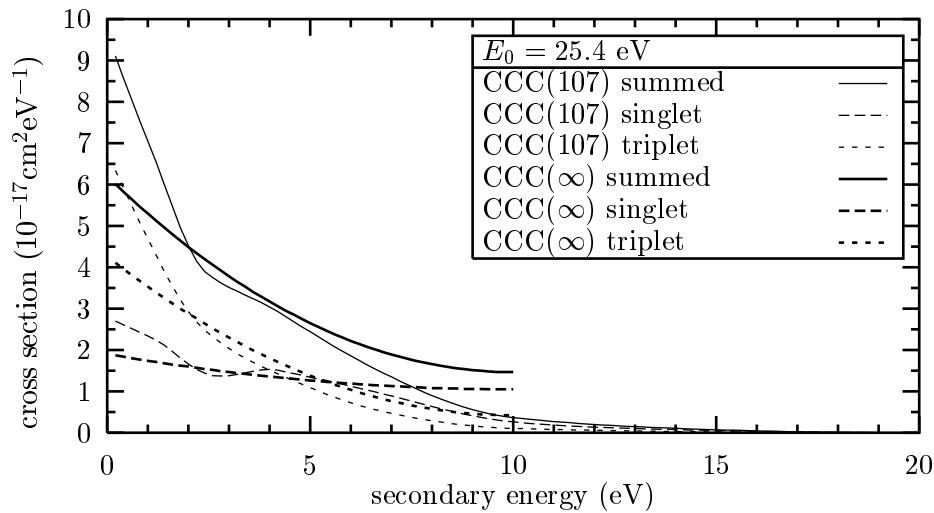}
\caption{The singly differential cross section for 25.4~eV
electron-impact ionization of the ground state of lithium calculated
using the 107-state CCC calculation. Also given are integral
preserving estimates labelled by CCC($\infty$), see text.
The singlet and triplet
cross sections include the spin weights.}
\label{25.4_sdcs}
\end{figure}

Firstly, in \fref{25.4_sdcs}, the SDCS are given. We see 
that the individual CCC(107) singlet and triplet SDCS have unphysical
oscillations. These are due to, we suspect, the SDCS($E/2$) being
substantial. Also given are the quadratic integral preserving estimates,
labelled as CCC($\infty$), whose SDCS($E/2$) are four times the
CCC(107) SDCS($E/2$). The label $\infty$ is used to suggest the result 
of a fully convergent close-coupling calculation. Note that at $E/2$
the convergence of CCC($\infty$) is to CCC(107) SDCS($E/2$), that is a 
quarter of the step size owing to convergence of CCC(107) amplitudes
to half the true amplitude at $E/2$. 

There is quite substantial
difference between  
the  CCC($\infty$) and CCC(107) curves, not only around $E/2$ but also 
at small secondary energies. While we do not know how accurate the
CCC($\infty$) estimates are we believe them to be more accurate than
the CCC(107) raw results and would use them to rescale the
magnitudes of any TDCS calculated for $E_B<E/2$.

\begin{figure}
\epsffile{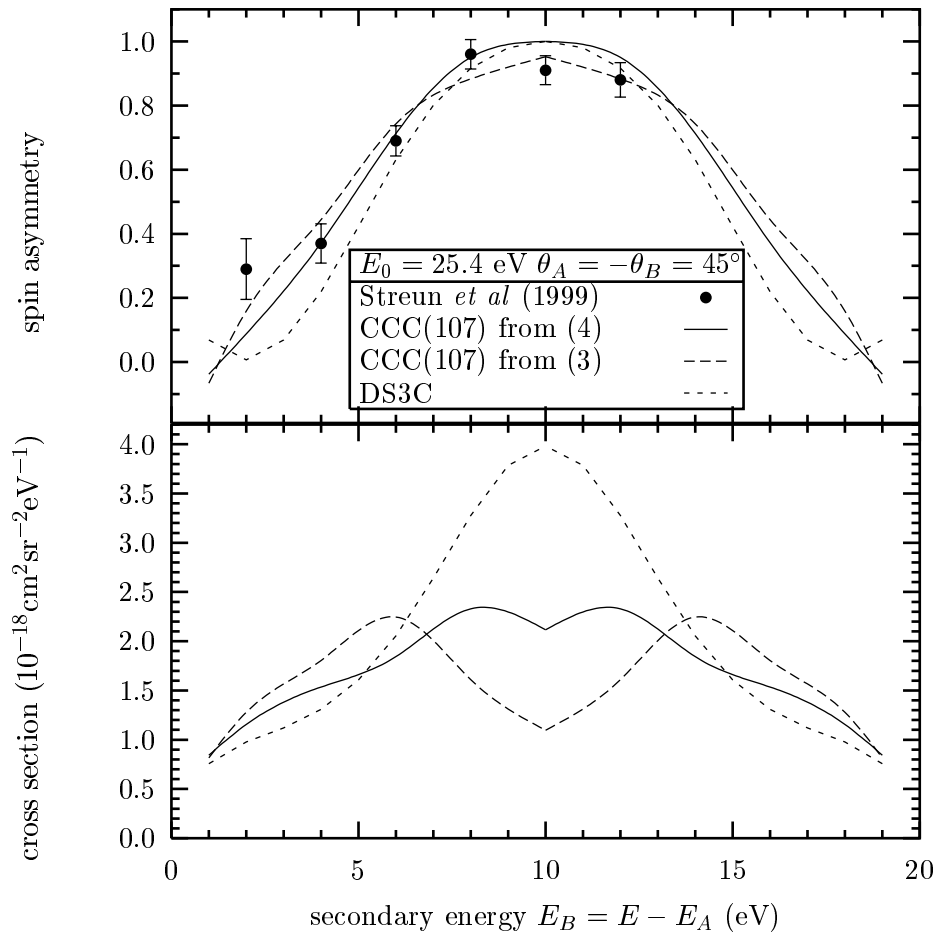}
\caption{The coplanar $\theta_A=-\theta_B=45^\circ$ triply
differential cross section and spin asymmetry for 25.4~eV 
electron-impact ionization of the ground state of lithium calculated
using the 107-state CCC calculation for both Eqs.\eref{old} and
\eref{new}. Also given are the DS3C results of \protect\citeasnoun{Streun99}.
}
\label{25.4_tdcs}
\end{figure}

\Fref{25.4_tdcs} gives the CCC(107) spin asymmetries and TDCS corresponding to
the experiment and calculations of \citeasnoun{Streun99}. Turning our
attention to the spin asymmetries we see very good agreement between
the CCC(107) calculation, using \eref{old} and \eref{new}, the DS3C
calculation, and the experiment. The agreement between the use of
\eref{new} and DS3C is a little better than the use of \eref{old}. This is
particularly evident around $E/2$ where both yield exactly unity, as would be
expected. The discrepancy with experiment in this case is due to
experimental uncertainties. One may ask what effect would rescaling
the CCC calculations according to the CCC($\infty$) estimates given in 
\fref{25.4_sdcs}. For the $E_B=2$~eV no change would result as here
the estimates and the raw results intersect. At $E_B=4$~eV the
estimates suggest a small increase in the triplet component and a small
decrease in the singlet component. This would result in a decrease in
the asymmetry (see \eref{asym}), and hence the DS3C calculation is
perhaps more accurate here, even though the relevant CCC result goes
through the midpoint of the error bar. At $E_B=6$~eV the estimates
predict a drop in the triplet cross section and therefore a further
small rise in the asymmetry. Lastly, for $E_B=8$~eV both the singlet
and triplet cross sections estimates are twice the raw results,
unaffecting the asymmetry.

The good agreement between the theories for the spin asymmetry is
lost when the corresponding TDCS is considered. The factor of two
difference at $E/2$ between the CCC(107) results evaluated using
\eref{old} and \eref{new} is expected. Thus there is no way for
\eref{old} to yield accurate SDCS in the vicinity of $E/2$. The
question is how accurate is the coherent combination \eref{new}?
While we are confident of the accuracy of the magnitude yielded by
\eref{new} at $E/2$, which is considerably lower than that predicted by the
DS3C theory, the cusp here looks somewhat unphysical. Given that detailed
convergence studies in a model problem show an ever increasing slope of the
CCC-calculated SDCS at $E/2$ \cite{B97l}, even the new form \eref{new}
may be unable to achieve 
accuracy in the SDCS in a neighbourhood of $E/2$, even while having
convergent SDCS at $E/2$. 

Another interesting feature that arises from the consideration of the
TDCS is the visible difference between the usage of \eref{old} and
\eref{new} for the asymmetric energy sharing below 5~eV. This is
surprising given the SDCS presented in \fref{25.4_sdcs}. The reason
for the difference is simple. If we write the ratio of the presented
SDCS(15)/SDCS(5)$=r^2$, then the ratio of the corresponding amplitudes 
is just $r$. Since
the combination \eref{old} sums cross sections, the
contributions past $E/2$ only significantly contribute to the
presented TDCS in the energy region 5 to 15~eV. However, the
combination \eref{new} has CCC amplitudes calculated at energies
greater than $E/2$ contributing visibly to the presented TDCS over a
much wider energy range.

Unfortunately, the difference between the two CCC-calculated TDCS is
somewhat academic since the underlying amplitudes have only converged
to an acceptable accuracy at the $E/2$ point. The difference between
the estimated and calculated SDCS in \fref{25.4_sdcs} is indicative of 
the lack of convergence (except at the $E/2$ point). Hence rescaling
the CCC-calculated TDCS utilising the SDCS estimates would
probably result in TDCS more accurate than those presented.

\section{Conclusions}
Spin-resolved electron-lithium ionization TDCS have been considered for 
the 54.4 and 25.4~eV incident energies. For both cases asymmetric
through to symmetric energy-sharing kinematics were considered.  At
the higher energy reasonable convergence in the spin asymmetries
has been achieved and good agreement with available
experiment obtained in all cases. For the equal
energy-case the combination of amplitudes suggested by
\citeasnoun{S99l}, see \eref{new}, yields more accurate spin
asymmetries. This indicates that obtaining convergent underlying CCC amplitudes
of correct symmetry is more difficult in the case of lithium than
for atomic hydrogen \cite{B99jpb}. The explicit imposition of the
required symmetry via \eref{new} allows for a faster rate of
convergence in observable phenomena.

At the lower energy the SDCS at $E/2$ is relatively more substantial
than at the higher energy and therefore the convergence in the
CCC ionization amplitudes has not been achieved generally. Numerically, it is
too difficult to adequately reproduce a step-function in the
underlying CCC amplitudes. Nevertheless, convergence of the $E/2$
amplitudes to half the true ones has been achieved to a reasonable
accuracy. The agreement with the measured and the DS3C-calculated spin 
asymmetries presented by \citeasnoun{Streun99} is very good. Thus the
spin asymmetry
experiment is unable to establish the relative accuracy of the CCC and 
DS3C theories. However, the DS3C and CCC
TDCS are very different, particularly at the equal energy-sharing
point where the CCC result is fully ab initio.
Accordingly, absolute determination of the ionization cross 
sections, preferably as a function of secondary energy would be very welcome.
\ack
We thank Jamal Berakdar for providing his calculations in electronic form.
Support of the Australian Research Council
and the Flinders University of South Australia is acknowledged.
We are also indebted to the South Australian
Centre for High Performance Computing and Communications.  

\section*{References}

\end{document}